\documentclass[12pt]{article}

\newcommand{\Dslash}{\not \!\! D}

\usepackage{amsmath,amssymb,amscd,amsbsy,amsgen,amsopn,amstext,
amsxtra}

\usepackage[dvips]{graphicx}

\begin{document}

\begin{flushright}
\end{flushright}

\vskip 0.5 truecm

\begin{center}
{\Large{\bf Quantum anomalies and some recent developments\footnote{Invited talk given at PAQFT08, November 27-November 29, 2008, Nanyang Technological University, Singapore}
 }}
\end{center}
\vskip .5 truecm
\centerline{\bf  Kazuo Fujikawa}
\vskip .4 truecm
\centerline {\it Institute of Quantum Science, College of 
Science and Technology}
\centerline {\it Nihon University, Chiyoda-ku, Tokyo 101-8308, 
Japan}
\vskip 0.5 truecm

\begin{abstract}
Some of the developments related to quantum anomalies and path integrals during the past 10 years are briefly discussed. The covered subjects include the issues related to the local counter term in the context of 2-dimensional path integral bosonization and the treatment of chiral anomaly and index theorem on the lattice. We also briefly comment on a recent analysis of the connection between the two-dimensional chiral anomalies and the four-dimensional black hole radiation.
\end{abstract}

\makeatletter
\@addtoreset{equation}{section}
\def\theequation{\thesection.\arabic{equation}}
\makeatother

\section{Introduction}

The modern quantum field theory formulated around 1947 by Tomonaga, Schwinger and  Feynman enormously widened the scope of quantum theory, and layed the foundation for the entire later developments of theoretical particle physics. The subject of quantum anomalies appeared immediately after the formulation of modern quantum field theory in 1949~\cite{bertlmann}, but it became a major subject only after 1969~\cite{bell, adler}.

In the present talk, I would like to discuss some of the developments in the subject during the past 10 years, whcih I found interesting. I concentrate on the  path 
integral formulation of quantum anomalies~\cite{fujikawa}.  
I first discuss the issue of local counter terms in defining quantum anomalies. In the conventional definition of quantum anomalies, those extra terms  which can be eliminated by local counter terms are classified as  the 
spurious anomalies and thus not regarded as genuine anomalies. 
This definition needs to be modified when one analyzes the path
integral bosonization, and in fact, I show that the naive local counter terms in the context of gauge theory are not local in the context of path integral bosonization.
This bosonization is also relevant in connection with the later discussion of black hole 
radiation and quantum anomalies.

I next comment on the chiral symmetry in lattice gauge theory
by concentrating on the notion of the index on the lattice, which enables us to define the chiral anomalies as the Jacobian 
in path integral formulation.
Several salient features of the lattice theory are also explained.
These features may turn out to be useful in the future analysis of gauge field theory in general. 

I finally comment on a rather surprising relation between the chiral anomaly in two-dimensions and the black hole radiation.
Here the ultra-local nature of the anomaly and the covariant form of chiral anomalies play a central role.

\section{Local counter terms and path integral bosonization}

The path integral bosonization means, for example, the free 
Abelian nosonization in two dimensions
\begin{eqnarray}
e^{iW(v_\mu)}&=&
\int{\cal D}\bar\psi{\cal D}\psi
\exp\left[i\int d^2x\left(\bar\psi i\gamma^\mu\partial_\mu\psi
+v_\mu\bar\psi\gamma^\mu\psi\right)\right]\nonumber\\
&=&\int{\cal D}\xi\exp\left[
i\int d^2x\left({1\over2}\partial^\mu\xi\partial_\mu\xi
-\frac{v_\mu}{\sqrt{\pi}}
\epsilon^{\mu\nu}\partial_\nu\xi\right)\right]
\end{eqnarray}
where $\xi$ is a real scalar field~\cite{roskies,gamboa-savari1,furuya}.
Other examples of bosonization include the correspondence between
the massive Thirring model and the sine-Gordon model
~\cite{coleman}, or the  non-Abelian free fermion and the the  non-linear $\sigma$
model (the WZW model)~\cite{witten}. All these cases are treated by path integrals \cite{bertlmann}.

We start with a theory which contains $U(1)$ gauge fields
\begin{eqnarray}
&&\int{\cal D}\bar\psi{\cal D}\psi
\exp\left(i\int d^2x\,{\cal L}\right)
=\int{\cal D}\bar\psi{\cal D}\psi\nonumber\\
&&\times
\exp\left\{i\int d^2x\left[\bar\psi i\gamma^\mu 
(\partial_\mu-iV_\mu-iA_\mu\gamma_5)\psi\right]\right\}.
\end{eqnarray}
For the chiral transformation
\begin{eqnarray}
\psi(x)\to\psi'(x)=e^{i\alpha(x)\gamma_5}\psi(x),\ \
\bar\psi(x)\to\bar\psi'(x)=\bar\psi(x)e^{i\alpha(x)\gamma_5}
\end{eqnarray}
the Jacobian is given by the master formula of quantum anomaly with $\epsilon^{10}=1$
\begin{eqnarray} 
\ln J_5(\alpha)={i\hbar\over\pi}\int d^2x\,\alpha(x) \left(\partial^\mu A_\mu+{1\over2}\epsilon^{\mu\nu}F_{\mu\nu}\right)
\end{eqnarray}
where we added $\hbar$ to emphasize that this is the one-loop 
effect, and 
\begin{eqnarray}
F_{\mu\nu}\equiv\partial_{\mu}V_{\nu}
-\partial_{\nu}V_{\mu}.
\end{eqnarray}
The anomaly evaluated above satisfies\\
1. Wess-Zumino integrability condition~\cite{wess}\\
2. Fermion number (vector current) conservation\\
3. $\epsilon_{\mu\nu}\bar{\psi}\gamma^\mu\psi
=\bar{\psi}\gamma_\nu\gamma_5\psi$\\
and thus suitable for bosonization.

If one follows the conventional wisdom, the first term  in the 
above Jacobian, $\partial^{\mu}A_{\mu}$, 
is eliminated by a {\em local counter term} $\frac{1}{2}A^{\mu}A_{\mu}$ (corresponding to the mass term of the gauge field $A_{\mu}$) and thus has no physical meaning. 
However, this term does not diverge in two-dimensional 
theory and, in fact, it plays a central role by giving the 
kinetic term for the boson field in  bosonization.

\subsection{  Anomalies and  bosonization}

We start with
\begin{eqnarray}
e^{iW(v_\mu)}=\int{\cal D}\bar\psi{\cal D}\psi
\exp\left[i\int d^2x\left(\bar\psi i\gamma^\mu\partial_\mu\psi
+v_\mu\bar\psi\gamma^\mu\psi\right)\right].
\end{eqnarray}
In this setting any {\it local\/} counter term should be 
expressed as a {\em local polynomial} in~$v_\mu$.
Now we observe that the vector field in two-dimensional 
space-time is 
decomposed into two arbitrary real functions~$\alpha$ 
and~$\beta$ as 
$v_\mu(x)=\partial_\mu\alpha(x)+\epsilon_{\mu\nu}
\partial^\nu\beta(x).$
We can then write 
\begin{eqnarray} 
e^{iW(v_\mu)}=\int{\cal D}\bar\psi{\cal D}\psi
\exp\left\{i\int d^2x\left[\bar\psi i\gamma^\mu 
(\partial_\mu-i\partial_\mu\alpha-i\partial_\mu\beta\gamma_5)\psi
\right]\right\}
\end{eqnarray}
by noting $\epsilon_{\mu\nu}\gamma^\mu=\gamma_\nu\gamma_5$.
 
 We extract 
the functions $\alpha$ and~$\beta$ as integrated Jacobians  associated with the transformations of integration variables 
$\psi$ and~$\bar\psi$. 
For infinitesimal transformations 
\begin{eqnarray}
&&\psi'(x)=\exp\left\{i\left[\delta\alpha(x)
+\delta\beta(x)\gamma_5\right]\right\}\psi(x),
\nonumber\\
&&\bar\psi'(x)=\bar\psi(x)
\exp\left\{i\left[-\delta\alpha(x)+\delta\beta(x)\gamma_5\right]
\right\},
\end{eqnarray}
we have the Jacobians $\ln J(\delta\alpha)=0$, and 
\begin{eqnarray}
&&\ln J(\delta\beta)=
{i\over\pi}\int d^2x\,\delta\beta(x) 
(\partial^\mu A_\mu+\epsilon^{\mu\nu}\partial_\mu V_\nu). 
\end{eqnarray} 
where $V_\mu=\partial_\mu\alpha$ and $A_\mu=\partial_\mu\beta$. 
Using these Jacobians, 
the $\alpha$ and $\beta$ dependences in the action
are extracted as
\begin{eqnarray}
e^{iW(v_\mu)}=\exp[i\Gamma(v_\mu)]\int{\cal D}\bar\psi
{\cal D}\psi \exp\left[i\int d^2x\,
(\bar\psi i\gamma^\mu\partial_\mu\psi) \right] 
\end{eqnarray}
 where $\Gamma(v_\mu)$ stands for the integrated Jacobian 
(or anomaly), and it has the form in Minkowski metric 
\begin{eqnarray} 
i\Gamma(v_\mu)={i\over\pi}\int d^2x\int_0^1ds\, 
\beta\partial^\mu(1-s)\partial_\mu\beta
={i\over\pi}\int d^2x \left(-{1\over2}\partial^\mu\beta
\partial_\mu\beta\right).
\end{eqnarray}
We can also write
\begin{eqnarray}
e^{iW(v_\mu)}=\int{\cal D}\xi\exp\left[
{i\over\pi}\int d^2x\left({1\over2}\partial^\mu\xi
\partial_\mu\xi\right)
\right]\exp[i\Gamma(v_\mu)]
\end{eqnarray}
since the absolute normalization of path integral does not 
matter in the definition of~$W(v_\mu)$. We next shift the variable $\xi\to\xi+\beta$, and  the ``translational invariance'' of the path integral measure ${\cal D}(\xi+\beta)={\cal D}\xi$ leads to
\begin{eqnarray}
e^{iW(v_\mu)}
&=&\int{\cal D}\xi\exp\left[
{i\over\pi}\int d^2x\left({1\over2}\partial^\mu\xi\partial_\mu\xi
+\partial^\mu\beta\partial_\mu\xi\right)\right]
\nonumber\\
&=&\int{\cal D}\xi\exp\left[
i\int d^2x\left({1\over2}\partial^\mu\xi\partial_\mu\xi
-\frac{v_\mu}{\sqrt{\pi}}
\epsilon^{\mu\nu}\partial_\nu\xi\right)\right].
\end{eqnarray}
In deriving the last line, we used 
$\partial^\mu\partial_\mu\beta=\epsilon^{\mu\nu}
\partial_\mu v_\nu$.

From (2.6) and (2.13) one sees that the theory of a free Dirac fermion 
$\psi$ and the theory of a free real Bose field $\xi$ define the  identical generating functional $W(v_\mu)$ of Green's functions. For example,
\begin{eqnarray} 
\langle T^*\bar\psi(x)\gamma^\mu\psi(x)\bar\psi(y)
\gamma^\nu\psi(y)\rangle
=\left({1\over\pi}\right)
\langle T^*\epsilon^{\mu\alpha}\partial_\alpha\xi(x)
\epsilon^{\nu\beta}\partial_\beta\xi(y)\rangle.
\end{eqnarray}

\subsection{  Local counter terms and bosonization}

The term~$\partial^\mu A_\mu$ in the Jacobian factor (2.9) plays a 
central role to give the 
kinetic term of the bosonic field~$\xi$. 
If one eliminates the term~$\partial^\mu A_\mu$ by a local 
counter term, the path integral 
bosonization as presented here does not work. 

The term 
$A_\mu^2/(2\pi)$, which is local in terms of the axial vector 
field~$A_\mu$, is  actually {\em not local} in the 
context of bosonization: The local counter term should be 
expressed as a local polynomial of the source field~$v_\mu$. 
The would-be counter  term is written in terms of~$v_\mu$ as~\cite{fuji-suzuki}  
\begin{eqnarray} 
{1\over2\pi}A_\mu^2&=&{1\over2\pi}(\partial_\mu\beta)^2
\nonumber\\
&=&-{1\over2\pi}\epsilon^{\mu\nu}\partial_\mu v_\nu {1\over\partial_\rho\partial^\rho}
\epsilon^{\alpha\beta}\partial_\alpha v_\beta 
\nonumber\\
&=&-{1\over8\pi}\epsilon^{\mu\nu}F_{\mu\nu}
{1\over\partial_\rho\partial^\rho}
\epsilon^{\alpha\beta}F_{\alpha\beta}
\end{eqnarray}
which is {\em not} local. It is thus not allowed to add this 
term as a counter term to the definition of the original 
partition function. This term, if added, 
modifies the physical contents of the original fermionic theory.

A {\em close analogue} appears in  the analysis of the 
Liouville action in the quantization of string theory~\cite{polyakov} 
 \begin{eqnarray} 
{\cal L}={1\over2}\sqrt{g}g^{\mu\nu}\partial_\mu X^a
\partial_\nu X^a
\end{eqnarray}
where the index $\mu$ of $x^\mu$ runs over $1$ and~$2$ and 
parameterizes the world sheet, and the index $a=1\sim d$ where 
$d$ stands for the dimension of the target space-time. 
In the conformal gauge 
$g_{\mu\nu}(x)=\rho(x)\eta_{\mu\nu}
=\exp[\sigma(x)]\eta_{\mu\nu}$, 
it is known that the (carefully defined) path integral measure in 
\begin{eqnarray} 
\int d\mu\exp\left[-\int d^2x\left( {1\over2}\sqrt{g}
g^{\mu\nu}\partial_\mu X^a \partial_\nu X^a\right)\right] 
\nonumber
\end{eqnarray}
gives rise to the Liouville action~\cite{fujikawa2}
\begin{eqnarray}
d\mu\to d\mu\,\exp\left[-{26-d\over48\pi}\int d^2x\left( {1\over2}\partial^\mu\sigma\partial_\mu\sigma+{1\over2}m^2e^{\sigma}
\right)\right]
\end{eqnarray}
when one extracts the Weyl freedom $\rho$ dependence from 
the action. 
The kinetic term of the Liouville action  appears to be 
eliminated by a suitable local counter term.
 But it is known that the kinetic term is non-local 
when written in a gauge condition other than 
the conformal gauge
\begin{eqnarray}
-{26-d\over96\pi}\int d^2x\,\sqrt{g}R\frac{1}{\Box}R 
\nonumber
\end{eqnarray}
where $R$ stands for the Riemann scalar curvature in 
two-dimensional space; in the conformal gauge one has 
$\sqrt{g}R=\partial^\mu\partial_\mu\sigma$. 

We thus conclude that 
{\em allowed local counter terms depend on physical situations}.

\section{ Lattice chiral symmetry and anomalies}

A siginificant development in the treatment of lattice chiral symmetry took place during the past 10 years. This is based on the so-called  Ginsparg-Wilson relation for the lattice Dirac operator  $D$~\cite{ginsparg}
\begin{eqnarray}
\gamma_{5}(\gamma_{5}D)+(\gamma_{5}D)\gamma_{5}=
2a(\gamma_{5}D)^{2}
\end{eqnarray}
with $a$ the lattice spacing. The explicit construction of $D$ free of species doublers was given~\cite{neuberger} and the correct index on the lattice has been established~\cite{hasenfratz}. The anomaly is then identified with the Jacobian~\cite{luscher} as in continuum theory.

We deal with a hermitian Dirac operator 
$H=a\gamma_{5}D=H^{\dagger}=aD^{\dagger}\gamma_{5}$
and write the relation (3.1) as 
\begin{eqnarray}
\gamma_{5}H+H\gamma_{5}=2H^{2}.
\end{eqnarray}
We also assume that the operator $H$ is local in the sense that 
it is analytic in the entire Brillouin zone.
One can confirm the relation
$\gamma_{5}H^{2}=H^{2}\gamma_{5}$.

The defining algebra (3.2) is written in various ways such as  
\begin{eqnarray}
&&\Gamma_{5}H+H\Gamma_{5}=0,\ \ (\Leftrightarrow \gamma_{5}\Dslash + \Dslash\gamma_{5}=0), \nonumber\\
&&\gamma_{5}H+H\hat{\gamma}_{5}=0,\nonumber\\
&&\hat{\gamma}^{2}_{5}=1,
\end{eqnarray}
 where
\begin{eqnarray}
\Gamma_{5}=\gamma_{5}-H,\ \ \
\hat{\gamma}_{5}=\gamma_{5}-2H.
\end{eqnarray}
We have 3 gamma matrices, $\gamma_{5},  \ \Gamma_{5},  \ 
\hat{\gamma}_{5}$, all of which agree for $a\rightarrow 0$.

We now examine the Euclidean action defined by
\begin{eqnarray}
S=\int d^{4}x\bar{\psi}D\psi\equiv\sum_{x,y}\bar{\psi}(x)D(x,y)
\psi(y)
\end{eqnarray}
which is invariant under
$\delta\psi=i\epsilon\hat{\gamma}_{5}\psi, \ \ 
\delta\bar{\psi}=\bar{\psi}i\epsilon\gamma_{5}$.
The chiral Jacobian for this $U(1)$ transformation becomes
\begin{eqnarray}
\ln J=-i\epsilon Tr(\hat{\gamma}_{5}-\gamma_{5})
=2i\epsilon Tr \Gamma_{5}=2i\epsilon (n_{+}-n_{-})
\end{eqnarray}
by using the index theorem on the lattice
\begin{eqnarray}
Tr \Gamma_{5}=n_{+}-n_{-}.
\end{eqnarray}

\subsection{Index relation on the lattice}
We start with the complete set of solutions
\begin{eqnarray}
H\varphi_{n}=\lambda_{n}\varphi_{n},\ \ \ 
(\varphi_{n},\varphi_{l})=\delta_{n,l}.
\end{eqnarray}
Then the index is defined as
\begin{eqnarray}
{\rm index} =n_{+}-n_{-},
\end{eqnarray}
where $n_{\pm}={\rm No.\  of\ zero\ modes}\ H\varphi_{n}=0,\ \ 
{\rm with} \ \ \gamma_{5}\varphi_{n}=\pm\varphi_{n}$, respectively.
In comparison, the continuum Atiyah-Singer index theorem reads
\begin{eqnarray}
\lim_{M\rightarrow\infty}Tr\gamma_{5}
\exp[-\frac{\Dslash^{2}}{M^{2}}]&=&n_{+}-n_{-}
=\nu \sim \int F\tilde{F}.
\end{eqnarray}

By using  the GW relation
\begin{eqnarray}
\Gamma_{5}H+\Gamma_{5}H=0, \ \
\end{eqnarray}
with $\Gamma_{5}=\gamma_{5}-H$, all the normalizable 
eigenstates $\phi_{n}$ of $\gamma_{5}D=H/a$ are categorized into 
the following 3 classes~\cite{fujikawa3}:\\
(i)\ $n_{\pm}$ (``zero modes''),
$H\phi_{n}=0, \ \ \gamma_{5}\phi_{n} = \pm \phi_{n}$\\
(ii)``paired states'' with $0 < |\lambda_{n}| < 1/a$,
\begin{eqnarray}
\frac{1}{a}H\phi_{n}= \lambda_{n}\phi_{n}, \ \ \ 
\frac{1}{a}H(\Gamma_{5}\phi_{n})
= - \lambda_{n}(\Gamma_{5}\phi_{n}).
\end{eqnarray}
(iii)\ $N_{\pm}$ (``highest states'', $\Gamma_{5}\phi_{n}=0$), 
\begin{eqnarray}
\frac{1}{a}H\phi_{n}= \pm \frac{1}{a}\phi_{n}, \ \ \
\gamma_{5}\phi_{n} = \pm \phi_{n},\ \ \ {\rm respectively}.
\end{eqnarray}

We thus obtain the index relation
\begin{eqnarray}
Tr\Gamma_{5}&\equiv& \sum_{n}(\phi_{n},\Gamma_{5}\phi_{n})
\nonumber\\
&=&\sum_{ \lambda_{n}=0}(\phi_{n},\Gamma_{5}\phi_{n})
+\sum_{0<|\lambda_{n}|<1/a}(\phi_{n},\Gamma_{5}\phi_{n})
+\sum_{|\lambda_{n}|=1/a}(\phi_{n},\Gamma_{5}\phi_{n})
\nonumber\\
&=&\sum_{\lambda_{n}=0}(\phi_{n},\Gamma_{5}\phi_{n})
=\sum_{\lambda_{n}=0}(\phi_{n},(\gamma_{5}-H)\phi_{n})\nonumber\\
&=&\sum_{\lambda_{n}=0}(\phi_{n},\gamma_{5}\phi_{n})
= n_{+} - n_{-} =  {\rm index}.
\end{eqnarray}

One can understand the physical meaning of $N_{\pm}$ states
with $\Gamma_{5}\phi_{n}=0$, whcih are peculiar with the present formulation, by noting 
\begin{eqnarray}
Tr \gamma_{5}&=&\sum_{\lambda_{n}=0}
(\phi_{n},\gamma_{5}\phi_{n})
+\sum_{\lambda_{n}\neq0}(\phi_{n},\gamma_{5}\phi_{n})
\nonumber\\
&=&n_{+}-n_{-}+\sum_{\lambda_{n}\neq0}a\lambda_{n}
\nonumber\\
&=&n_{+}-n_{-}+N_{+}-N_{-}=0
\end{eqnarray}
namely,{\em chirality is always balanced} for a finite system~\cite{chiu,fujikawa3}
\begin{eqnarray}
n_{+}+N_{+}=n_{-}+N_{-}.
\end{eqnarray}
The reason we get a non-vanishing index for 
$Tr\Gamma_{5}=n_{+}-n_{-}$
is that $N_{\pm}$ states are projected out in $Tr\Gamma_{5}$.
The presence of $N_{\pm}$ is essential for a consistent definition of index for a finite system.

The {\em chiral} gauge theory is defined by
\begin{eqnarray}
\int {\cal D}\bar{\psi}_{L}{\cal D}\psi_{L}
\exp[\int\bar{\psi}P_{+}D
\hat{P}_{-}\psi]
\end{eqnarray}
with
$P_{+}=\frac{1}{2}(1+\gamma_{5}),\ \ 
\hat{P}_{-}=\frac{1}{2}(1-\hat{\gamma}_{5})$ since 
$D=P_{+}D\hat{P}_{-}+P_{-}D\hat{P}_{+}$.
For chiral gauge theory, fermion number transformation
\begin{eqnarray}
\psi^{\prime}=e^{i\alpha}\psi, \ \ \ \ 
\bar{\psi}^{\prime}=\bar{\psi}e^{-i\alpha}
\end{eqnarray}
gives the Jacobian
\begin{eqnarray}
&&\ln J=-i\alpha Tr(\hat{P}_{-}-P_{+})=i\alpha
Tr(\Gamma_{5})
=i\alpha(n_{+}-n_{-}).
\end{eqnarray}
Namely, the {\em covariant} fermion number anomaly~\cite{'t hooft} is automatically built in.

A {\em new} feature of lattice chiral gauge theory compared to continuum theory is that we can define the generators of chiral
gauge symmetry 
\begin{eqnarray}
P_{\pm}T^{a}, \ \ \ \ [P_{\pm}T^{a},P_{\pm}T^{b}]=if^{abc}
P_{\pm}T^{c}
\end{eqnarray}
in continuum theory, but on the lattice we basically start with a vector-like theory and the chiral theory is defined by a projection
\begin{eqnarray}
\bar{\psi}D(U)\psi\ \ \Rightarrow \ \ 
\bar{\psi}P_{+}D(U)\hat{P}_{+}(U)\psi.
\end{eqnarray}

\section{ Black hole radiation from anomalies}

Robinson and Wilczek~\cite{robinson} have shown that the black hole radiation is understood as a result of chiral anomalies in two-dimensions. Note that  both of quantum anomalies  and Hawking radiation are of the order $\sim O(\hbar)$. We here briefly comment on this interesting idea. (The use of the conformal anomaly in two-dimensional gravitational theory is well-known~\cite{christensen}.)

The four dimensional action for a real scalar field $S_{(O)}(\phi, g^{\mu\nu}_{(4)})$, which is defined at a distance away from the black hole, is described by an effective two-dimensional theory
$S_{(H)}(\phi,g^{\mu\nu}_{(2)},A_{\mu},\Phi)$  near the horizon
after expanding $\phi$ in terms of spherical harmonics~\cite{robinson, iso}. 
Here $A_{\mu}$ and $\Phi$ are respectively two dimensional effective gauge field and dilaton field arising from $g^{\mu\nu}_{(4)}$.
We consider the Schwarzschild black hole and separate the region
outside the horizon into two regions, one near the black hole $r_{+}\leq
r< r_{+}+\epsilon$ (denoted by H) and the other away from the horizon 
$r_{+}+\epsilon\leq r$ (denoted by O) where $r_{+}$ stands for the black hole radius and $\epsilon$ is a finite but small quantity.

In the region away from the horizon, we have the Ward identity for general coordinate transformation
\begin{eqnarray}
\nabla_{\nu}T^{\mu\nu}_{(4)}=0 
\end{eqnarray}
to be consistent with the Einstein equation and the Bianchi identity. We thus have 
\begin{eqnarray}
\partial_{r}T_{t(O)}^{r}(r)=0,
\end{eqnarray}
where we defined $T_{t(O)}^{r}(r)=\int d\Omega_{(2)}r^{2}T_{t(4)}^{r}$, and thus $T_{t(O)}^{r}={\rm constant}$. To be precise, our energy-momentum tensor is defined by $T^{\mu\nu}_{(4)}(t,r)=\langle 0|
\hat{T}^{\mu\nu}_{(4)}|0\rangle$ in the vacuum specified by the 
Schwarzschild solution.
In the vicinity of the black hole, which is denoted by $H$, one can define an effective two-dimensional space-time described by $t$ and $r$ when one integrates over the angular freedom.
We then have the effectively two-dimensional relation
\begin{eqnarray}
\nabla_{\mu}T_{\nu(H)}^{\mu}(r)-\frac{\partial_{\nu}\Phi}{\sqrt{-g}}\frac{\delta S}{\delta \Phi}={\cal A}_{\nu},\ \ \ \mu,\nu=(t,r)
\end{eqnarray}
where we define $T_{\nu(H)}^{\mu}(r)$ as the energy-momentum 
tensor corresponding to the (right-handed) chiral freedom which is out-going from the horizon and ${\cal A}_{\nu}$ stands for the two-dimensional covariant gravitational anomaly~\cite{iso, banerjee, umetsu} for a chiral theory. Here one may use the "fermionization", the inverse of the bosonization. The other (left-handed) chiral fluctuation falling into the black hole is tentatively neglected.
The $\mu=t$ component of the Ward identity (4.3) becomes
\begin{eqnarray}
\partial_{r}T_{t(H)}^{r}(r)=\partial_{r}N_{t}^{r}(r)\nonumber
\end{eqnarray}
with the {\em covariant} gravitational anomaly~\cite{alvarez-gaume,fuji-tomita-yasuda}
\begin{eqnarray}
&&{\cal A}_{\nu}=\frac{1}{96\pi\sqrt{-g}}\epsilon_{\mu\nu}\partial^{\nu}R=\partial_{r}N_{\mu}^{r},\nonumber\\
&& N_{t}^{r}=\frac{ff^{\prime\prime}-(f^{\prime})^{2}/2}{96\pi},\ \ \ \ N_{r}^{r}=0\nonumber
\end{eqnarray}
where $f(r)=1-2M/r$.
We thus have
\begin{eqnarray}
\partial_{r}(T_{t(H)}^{r}(r)-N_{t}^{r}(r))=0.
\end{eqnarray}

At the boundary $r=r_{+}+\epsilon$ of two regions, we impose
\begin{eqnarray}
T_{t(O)}^{r}(r_{+}+\epsilon)=T_{t(H)}^{r}(r_{+}+\epsilon)-N_{t}^{r}(r_{+}+\epsilon)
\end{eqnarray}
which is consistent with continuity equation, i.e., the energy-momentum conservation. Also the covariance property of the both-hand sides matches. We interprete the second anomaly term on the right-hand side of (4.5) as the effective energy-momentum flux generated by the matter component which is in-going into the black hole; the second term (the effect of the in-going mode) vanishes 
for the formal limit $r_{+}+\epsilon \rightarrow \infty$, which is the condition of the so-called Unruh vacuum.
We now examine the small $\epsilon$ limit of this relation and impose the regularity condition $T_{t(H)}^{r}(r_{+})=0$ for the component escaping from the horizon. We thus obtain the ordinary result of black hole radiation~\cite{robinson, iso}
\begin{eqnarray}
T_{t(O)}^{r}(r)=T_{t(O)}^{r}(r_{+})=\frac{\pi}{12\beta^{2}}
\end{eqnarray}
since $T_{t(O)}^{r}(r)={\rm constant}$. Here
we defined the surface gravity
\begin{eqnarray}
&&\kappa = \frac{2\pi}{\beta}=\frac{1}{2}f^{\prime}(r_{+}).
\end{eqnarray}
 We admitted a certain asymmetry between the in-going and out-going components of the energy-momentum fluctuation in the balck hole vacuum by imposing different boundary conditions.
Our treatment slightly differs from the original treatments~\cite{robinson, iso} in that we defined all the anomalies in the covariant form which is naturally done in the path integral formualtion on the basis of the transformation of path integral variables and the evaluation of the resulting Jacobians~\cite{fujikawa}. The ultra-local property of the anomaly, namely, the fact that the anomaly is defined in any small but finite space-time region~\cite{fujikawa}, is also important in considering the small $\epsilon$ limit.

It is known that this consideration works also for Kerr and Reissner-Nordstroem black holes~\cite{iso, banerjee, umetsu}.

\end{document}